\documentclass[12pt]{article}
\usepackage{amssymb}
\usepackage{graphicx}
\usepackage{setspace}
\usepackage{color}
\newcommand{\n}{\nonumber \\}

\newcommand {\be}{\begin{eqnarray}}
\newcommand{\ee}{\end{eqnarray}}
\def\n{\nonumber \\}
\newcommand{\ltsim}{\protect\raisebox{-0.5ex}{$\:\stackrel{\textstyle <}{\sim}\:$}}

\begin{document}

\begin{flushright}
KEK-TH-1872 \\ LAPTH-064/15  
\end{flushright}
\vspace*{10mm}


\begin{center}
{\LARGE\bf
\begin{spacing}{1.2}
Dynamical fine-tuning of  initial conditions \\ for small field inflations
\end{spacing}
}
\vspace*{20mm}
{\large
Satoshi Iso${}^{\; a,b}$, Kazunori Kohri${}^{\; a,b}$ and Kengo Shimada${}^{\; c}$
}
\vspace{10mm}

{${}^a$\sl\small KEK Theory Center, High Energy Accelerator Research Organization (KEK),\\ }
{${}^b$\sl\small Graduate University for Advanced Studies (SOKENDAI), \\ Tsukuba, Ibaraki 305-0801, Japan \\}
\vspace{8pt}

{${}^c$\sl\small  LAPTh, Universit\'e de Savoie, CNRS, B.P. 110, F-74941 Annecy-le-Vieux, France \\ }

\vspace{8pt}
e-mails: {\small \it satoshi.iso(at)kek.jp, kohri(at)post.kek.jp , kengo.shimada(at)lapth.cnrs.fr }
\vspace{8pt}

\end{center}

\begin{abstract} 
Small-field inflation (SFI) is widely considered to be unnatural
  because an extreme fine-tuning of the initial condition
is necessary for sufficiently large e-folding. In this paper, we show that
the unnaturally-looking initial condition can be 
dynamically realised without any fine-tuning 
if the SFI occurs after  rapid oscillations of the inflaton field and particle creations by preheating. 
In fact, if the inflaton field $\phi$ is coupled to another scalar field $\chi$ through the interaction
$g^2 \chi^2 \phi^2$ and the vacuum energy during the small field inflation is given by
$\lambda M^4$, the initial value can be dynamically set at 
$(\sqrt{\lambda}/g) M^2/M_{\rm pl}$, which is much smaller than the typical scale 
of the potential $M.$ This solves the initial condition problem in the new inflation model
or some classes of the hilltop inflation models.
\\
\end{abstract}


\newpage
\section{ Introduction} 
\label{sec:introduction}
The recent observation of the precise CMB data by the Planck satellite\cite{Ade:2013zuv} 
gives an upper bound of the tensor to scalar ratio $r=16 \epsilon=\Delta_T^2/\Delta_S^2<0.12$ with 95 \% confidence level. Here the amplitudes of the tensor and the scalar 
fluctuations (in dimensionless form) are given by 
\begin{eqnarray}
&& \Delta_T^2(k) \equiv  \frac{k^3}{2\pi^2} {\cal P}_T (k)=\frac{2 \rho}{3 \pi^2 M_{\rm pl}^4},  \ \ \ 
 \Delta_S^2 (k) \equiv  \frac{k^3}{2\pi^2} {\cal P}_\zeta (k)= \frac{\rho}{24 \pi^2 M_{\rm pl}^4 \epsilon}   \  .
\end{eqnarray}
$\rho$ is the  energy density of the universe and related to the Hubble parameter 
by the Einstein equation
$H^2=\rho/3M_{\rm pl}^2.$ 
 $M_{\rm pl}=2.4 \times 10^{18}$ GeV is the reduced Planck scale.
The scalar fluctuations (curvature perturbations) \cite{Ade:2013zuv} 
 are given by $\Delta_S^2 =2.215 \times 10^{-9} $
at the pivot scale $k_{\rm CMB}=0.05$ Mpc$^{-1}$. 
This constrains the energy scale of the primordial inflation 
$\rho^{1/4}< 1.9 \times10^{16}$ GeV. 

Large field inflation models often predict larger values of $r$ than the observed value,
and  
it  gives a chance of revival to  small field inflations (SFI). 
 However, SFI has been known to have some drawbacks.  
First, SFI often  predicts smaller spectral index of the scalar perturbations
compared to the observed value $n_s \sim 0.96$. The problem can be solved by 
introducing a tilt (a linear type potential) in the inflaton potential (see, for example, 
\cite{EJChun}\cite{Nakayama:2012dw}\cite{Takahashi:2013cxa}\cite{IKS}). 
Another long-standing problem of SFI is that it
requires very fine-tuning of the initial condition. In order to explain the sufficiently large
e-folding, it is necessary to put the initial value of the SFI very close to the top of the 
potential in the case of the Coleman-Weinberg type inflation.

The purpose of the paper is to show that such an unnaturally-looking initial condition
can be dynamically fixed by using the preheating mechanism \cite{preheating,preheating2} without 
introducing any fine-tunings
of the initial condition. We only require that the SFI follows rapid oscillation of 
the inflaton field which had produced large number of particles and modified 
the inflaton potential so that the inflaton field is  trapped near the origin.
The mechanism is similar to the moduli trapping mechanism discussed  in \cite{beauty}, and
 also discussed previously in the context of the SFI \cite{Kofman:1995fi}\cite{Felder:2000sf}.
 In this paper, we revisit the problem and 
 show  that the initial value of the SFI is dynamically set at a sufficiently 
close point near the origin; $\phi_{\rm ini} \sim (\sqrt{\lambda} /g) M^2/M_{\rm pl} \ll M$ 
where the vacuum energy there is given by $\sim \lambda M^4$ and $g$ is the strength of 
interaction $g^2 \phi^2 \chi^2$ between the inflaton $\phi$ and another scalar field
$\chi$. 

The organisation of the paper is as follows.
In sec \ref{sec:SFI}, the fine-tuning problem of the initial condition
for the SFI is explained. In sec \ref{sec:LFI}, we briefly comment on
 the large field inflation in the Coleman-Weinberg potential.  
In sec \ref{sec:preheat}, the preheating mechanism is discussed, and we 
obtain the conditions for the broad parametric resonance to occur.
In sec \ref{sec:backreaction}, we study the effect of
 the created particles by preheating
on the coherent motion of the inflaton,  and show when the broad parametric resonance ends.
In sec \ref{sec:trap}, we show that the created particles can trap the inflaton field
near the top of the hill of the potential. Finally in sec \ref{sec:dynamicaltune},
we show that the unnaturally-looking initial condition of the SFI can be 
dynamically set without any fine-tuning, and summarize in sec \ref{sec:summary}.
\\
\section{ Small field inflation and the fine-tuning problem}
\label{sec:SFI}
\begin{figure}[t]
 \begin{center}
\includegraphics[width=0.8 \linewidth,bb=0 0 720 540]{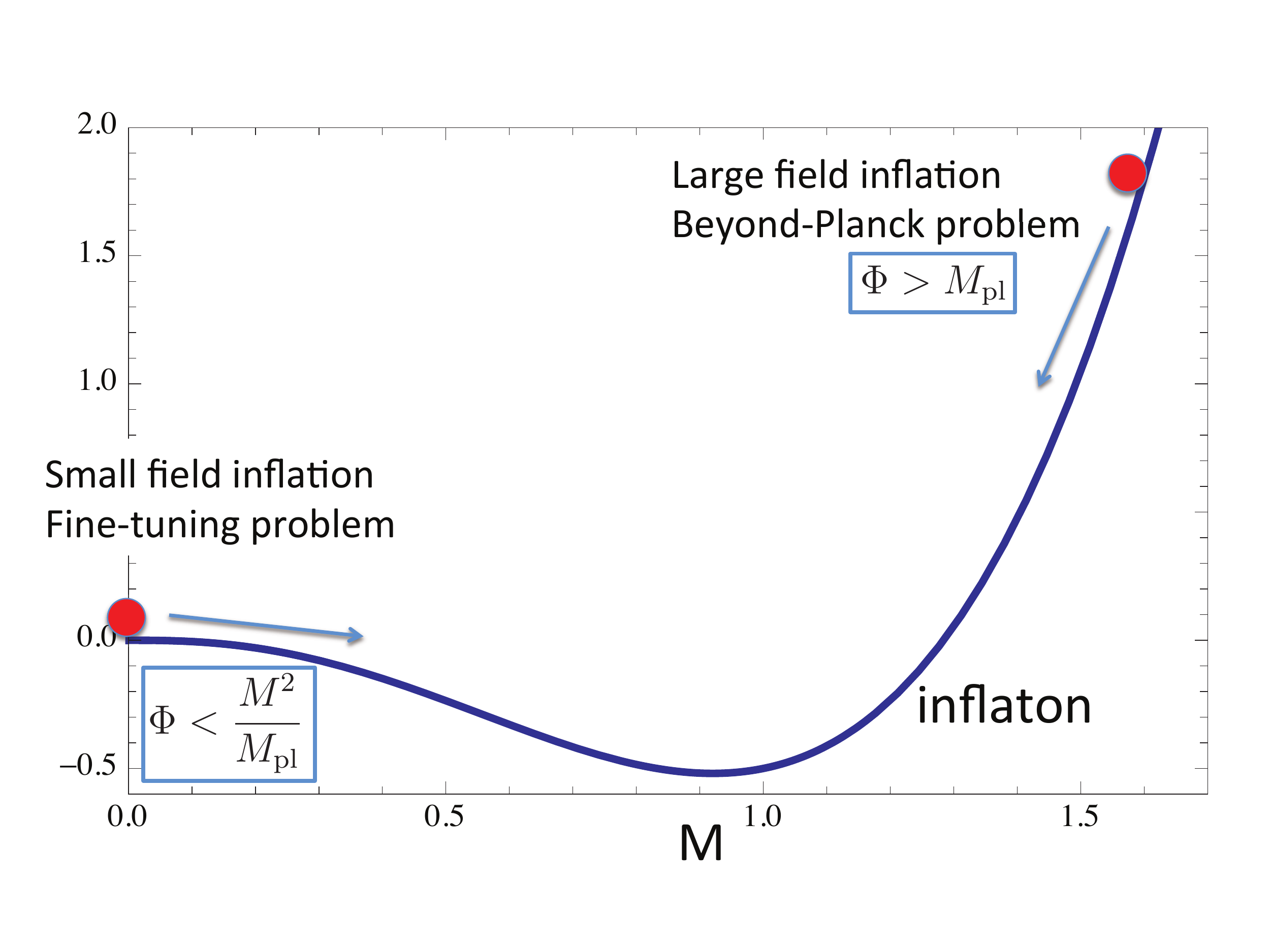}
   \caption{Coleman-Weinberg potential has the true minimum at $\phi=M$. It is flat at 
   the origin $\phi=0$. Two possibilities of inflationary scenario exist, the large field inflation (chaotic inflation) and the small field inflation (new inflation). The LFI has the beyond-Planck scale problem
   while the SFI has the fine-tuning problem of the initial condition.}
   \label{fig:CW}
 \end{center}
\end{figure}

We particularly consider the CW type  potential \cite{CW} shown in figure \ref{fig:CW};
\be
V(\phi) = \frac{\lambda \phi^4}{4} \left( \ln \frac{\phi^2}{M^2} -\frac{1}{2} \right) + V_0
\label{CWpotential}
\ee
where $V_0= \lambda M^4/8.$ 
The minimum of the  potential is given at $\phi=M$.
In this paper, we assume $M \ll M_{\rm pl}$.

Such potentials are studied in various situations, in particular, in the analysis 
of the radiative symmetry breaking of gauge theories, 
such as the grand unified theories (GUT) or extensions of the standard model (SM).
Especially, it has attracted renewed interest recently as a simple model 
of physics beyond the SM satisfying the experimental constraints imposed 
on the BSM by the LHC and flavor experiments. 
In \cite{IOO,IO}, based on the idea of \cite{naturalness}, 
we proposed a minimal extension of the SM 
by introducing $B-L$ (Baryon number minus Lepton number) 
$U(1)$ gauge field $Z'$, an additional scalar $\phi$ whose vacuum
expectation value (VEV) breaks the $B-L$ gauge symmetry, and the right handed neutrinos
which cancel the gauge anomaly of the $B-L$ symmetry.
We show that the electroweak (EW) gauge symmetry breaking 
is triggered by the $B-L$ breaking, which is radiatively broken by the CW mechanism.
The model is a minimal extension of the SM in which radiative symmetry breaking
can generate the EW scale. One of our motivations of the present analysis is 
to investigate the cosmological possibility of the model, but in order to 
make our discussions as general as possible, we do not use the specific
numerical values of the coupling constants in the following.

We make use of 
 the  scalar field $\phi$ as an inflaton field.
Since  $\phi$ has the CW type potential as in the Fig. \ref{fig:CW},
 two types of inflations are possible:  the large field inflation (LFI) and the small field
inflation (SFI).  
The large field type  can be  regarded as  the chaotic inflation
with a quartic potential.
On the other hand, the small field CW inflation was studied in the early
eighties in the non-supersymmetric GUT
models\cite{Linde:1981mu,Albrecht:1982wi,Shafi:1983bd}, and 
called the new inflation. In order to realize the SFI, 
it is often assumed that 
the inflaton field is trapped at the origin due to thermal corrections
to the effective potential generated in the reheating of the LFI.
When the fluctuations of the field are dominated by the vacuum
energy at $\phi=0$, the SFI occurs.
Then the radiation generated so far rapidly dilute,  and  the inflaton field $\phi$
rolls down  to the true minimum at $\phi=M$.  
The mechanism works  when the reheating occurs 
by perturbative decay of inflaton. 
Since the decay process produces relativistic particles, the modification of the inflaton potential is not always sufficient to lift the true minimum to trap  the field around the origin. 
Also, in order to explain the CMB fluctuations, 
the inflaton field must start from a very small value $\phi \ll M$  near the top
of the potential. In the following we explain how much fine-tuning is necessary.

First we calculate the slow roll parameters to estimate the necessary initial condition
 in the SFI. 
 Taking
derivatives with respect to $\phi$, we have 
\be V' = \lambda \phi^3 \ln
\frac{\phi^2}{M^2} , \ \ \  V'' = \lambda \phi^2 \left( 2 + 3 \ln
  \frac{\phi^2}{M^2} \right) .
\ee
Mass of the scalar at the minimum is given by $m_\phi^2=V''(M)=2 \lambda M^2.$
For small values of fields $(\phi < M)$, the slow roll parameters are calculated to be
\begin{eqnarray}
\epsilon &=& \frac{M_{\rm Pl}^2}{2} \left(\frac{V'}{V} \right)^2
 \approx 32 \left( \frac{M_{\rm Pl}}{M} \right)^2 
\left( \frac{\phi}{M} \right)^6 \left( \ln \frac{\phi^2}{M^2} \right)^2 \ , 
\\
 \eta &=& M_{\rm Pl}^2 \left( \frac{V''}{V} \right) \approx 
24 \left( \frac{M_{\rm Pl}}{M} \right)^2
\left( \frac{\phi}{M} \right)^2  \ln \frac{\phi^2}{M^2}  \  .
\label{eta}
\end{eqnarray}
Here we used $V \approx V_0$ in the region $\phi \ll M.$
The slow roll conditions $\epsilon, |\eta| <1$  require that
the field value $\phi$ during the SFI
must be extremely smaller than $M$; thus the relation $\epsilon \ll |\eta|$
 follows. Inflation stops at  $|\eta|=1$ where the slow roll condition is violated.
 
Eq. (\ref{eta}) can be approximately solved as
\be
\phi&\approx& \sqrt{\frac{|\eta|}{24}  
 \ln^{-1} \left( \frac{24 M_{Pl}^2}{|\eta|M^2} \right) }\frac{M^2}{M_{\rm pl}} .
 \label{fieldvalue}
\ee
It requires  that, in order to satisfy the 
the slow roll condition  with sufficiently large e-folding,  
 the inflation must start from the very small initial condition,
\be
\phi_{\rm ini} \sim (10^{-3}|\eta|)^{1/2}  \frac{M^2}{M_{\rm pl}} \ll M  .
\label{initialcondition}
\ee
This is the infamous fine-tuning problem of the initial condition of the SFI.  
In deriving the coefficient numerically, 
we inserted $M=10^{10}$ GeV and $|\eta|=0.02$  but 
the coefficient $10^{-3}$ does not depend so much on the details of these values.
For these values, the initial value needs to be 
$\phi_{\rm ini} \sim 10^{-11} M.$ Since $M$ is the typical scale of the potential (the 
energy scale at the minimum of the CW potential), such small coefficient $10^{-11}$
seems very unnatural as the initial condition. This is one of the reasons that
large field inflations, such as the chaotic ones, are  more favoured than the SFI.

 The slow roll parameter $\epsilon$ is much smaller than $|\eta|$ and given by
\be
\epsilon =  \frac{|\eta|^3}{432 \ \ln(24 M_{pl}^2/|\eta|M^2) }  
\left(\frac{M}{M_{Pl}} \right)^4 \sim 
4 \times10^{-5} |\eta|^3  \left(\frac{M}{M_{Pl}} \right)^4  \ll 1 \  .
\ee
In order to make the amplitude of the scalar perturbation $\Delta_S^2$
consistent with the Planck data \cite{Ade:2013zuv}, $\Delta_S^2 =
2.215 \times 10^{-9}$, 
the quartic coupling of the CW potential $\lambda$ must be extremely small
$\lambda \sim 10^{-15}$. 
In the following sections, we see that
the smallness of the  coupling becomes important to generate  rapid particle creations 
during coherent oscillations of the inflaton field.

Here  we comment on the issue of the smallness of the spectral 
index $n_s$.  
Those who are more interested in the initial condition problem can skip this paragraph.
The e-folding number $N$ of the SFI is related to the slow roll parameter $\eta$ as
\be
N &=& \frac{1}{M_{\rm Pl}^2} \int_{\phi_{end}}^{\phi} \frac{V}{V'} d \phi
\approx \frac{3 }{2  }
\left( \frac{1}{|\eta|} -  \frac{1}{|\eta_{end}|} \right) \  .
\ee
By putting $|\eta_{end}|=1$, we have   $\eta=-1/(2N/3+1).$
Since $\epsilon \ll |\eta|$, the spectral index  of the scalar perturbation is given by
$n_s=1-6\epsilon+2 \eta \sim 1+2 \eta$. 
Hence $n_s\sim 0.96$ \cite{Ade:2013zuv} requires an e-folding number 
$N=3/(1-n_s)-3/2=73.5$. 
On the other hand, 
the e-folding number at the pivot scale of CMB is given by
\begin{eqnarray}
   \label{eq:Ncond}
   N_{\rm CMB} &=& 61 + \frac23 \ln  \left(\frac{V_0^{1/4}}{10^{16} \mbox{GeV}} \right)
   + \frac13 \ln \left( \frac{T_R}{10^{16} \mbox{GeV}} \right),
   \label{CMBefolding}
\end{eqnarray}
where we assumed that there was an epoch of the inflaton field's
oscillation induced by its mass term after the inflation and then  the radiation
dominated epoch continues until the matter-radiation equality epoch. 
The smallness of the vacuum energy $V_0^{1/4} \sim 10^{-4} M \ll
M_{Pl}$ suggests a smaller e-folding number than $61$, which is inconsistent with
the above large e-folding number $N=73.5$.  
Various resolutions of the inconsistency have been proposed 
\cite{EJChun}\cite{Nakayama:2012dw}\cite{Takahashi:2013cxa}.
In our previous letter \cite{IKS}, we proposed another possibility that a
linear potential  is generated by the chiral condensates of quarks.

In the rest of the paper, we solve the fine-tuning problem of the initial condition given in (\ref{initialcondition}) by using the dynamics of preheating during the rapid oscillation of 
inflaton field before the SFI starts.

\section{Large field inflation and beyond-Planck problem}
\label{sec:LFI}
Before trying to solve  the initial condition problem of the SFI, we
remind the readers of the beyond-Planck scale problem in the LFI. 
If the inflation has the potential of the CW type in Figure \ref{fig:CW}, it is natural to think
that in the early universe
a coherent motion of the inflaton starts from a field value with $\phi \gg M.$ 
Then the potential can be approximated by the quartic one 
$V=\lambda \phi^4/2.$ 
The slow roll parameters in the region $\phi \gg M$ are given by
\be
\epsilon &=& \frac{M_{\rm Pl}^2}{2} \left(\frac{V'}{V} \right)^2
 \approx 8 \left(\frac{M_{\rm Pl}}{\phi} \right)^2  \\
\eta &=& M_{\rm Pl} \left( \frac{V''}{V} \right) \approx 
12 \left(\frac{M_{\rm Pl}}{\phi} \right)^2 = 3 \epsilon/2.
\ee
If the field value is much larger than the 
Planck scale $M_{\rm pl}$, the slow roll conditions are satisfied and the LFI occurs. 
The spectral index for the scalar perturbation is given by
$n_s = 1- 6 \epsilon + 2 \eta = 1-3 \epsilon .$
Hence, in order to explain the CMB data $n_s =0.9603 \pm 0.0073$, 
we need  $\epsilon \sim 0.013$, and accordingly 
the field value at the pivot scale $\phi \sim 25 M_{\rm pl}$ is necessary.
The tensor-scalar ratio is predicted as 
\be
r=16 \epsilon \sim 0.208,
\label{TSR-LFI}
\ee
and the e-folding is given by
\be
N &=& 
\int_{\phi_{end}}^{\phi} \frac{V}{M_{\rm Pl}^2 V'}d \phi
= \frac{1}{\sqrt{2} M_{\rm Pl}} \int_{\phi_{end}}^{\phi}
\frac{d \phi}{\sqrt{\epsilon}} 
= \frac{1}{8 M_{\rm Pl}^2} (\phi^2 -\phi_{end}^2)
\sim \frac{1}{\epsilon} -1
\label{LFinfN}
\ee
where we put $\epsilon_{end}=1$.

Two drawbacks are known in LFI. First it predicts large
 tensor to scalar ratio (\ref{TSR-LFI}) which seems to be inconsistent
with the Planck and BICEP2/Keck Array observations \cite{Ade:2013zuv,Array:2015xqh}. 
Second the large field value $\phi \gg M_{\rm pl}$  may invalidate
the analysis of the inflaton potential within renormalizable field theories, 
and higher mass-dimensional terms cannot be excluded.
 They are  common problems of the LFI.
We call the second one the beyond-Planck-scale problem in this paper.

In the following, we show that, in the CW type potential, 
the SFI naturally follows the LFI.  
Hence,  the CMB fluctuations are generated during the SFI, and the first problem is absent. 
Furthermore, as we show later, in order to  solve the initial
condition problem of the SFI (\ref{initialcondition}), 
it is sufficient to require that, before the SFI occurs,  
 the field starts from somewhere in an intermediate region $M< \phi< M_{\rm pl}.$
Hence the beyond-Planck-scale problem is also absent. 

\section{Preheating: broad parametric resonance}
\label{sec:preheat}
We consider a coupled system of two  scalar fields, an inflaton field $\phi$ and another 
scalar field $\chi$. The potential of the model is
\be
V(\phi, \chi) =V(\phi) + \frac{g^2}{2} \phi^2 \chi^2 \ ,
\ee
where $V(\phi)$ is given in (\ref{CWpotential}).
The model can be considered as a toy model of the 
classically conformal $B-L$ extended standard model \cite{IOO}.
The SM singlet scalar whose VEV breaks the $B-L$ gauge symmetry
 plays the role of the inflaton $\phi$, 
and the scalar field $\chi$ corresponds to the  $B-L$  $U(1)$ gauge field $Z'$. 
Hence the coupling $g$ represents  the $B-L$ gauge coupling $g_{B-L}$.
In the model \cite{IOO}, $Z'$ gauge boson is  coupled to the SM particles and decays into them.
In the present paper, we briefly comment on the effects of the $\chi$ decay
and leave detailed (numerical) analysis for future investigations.

The strengths of the two coupling constants, $\lambda$ and $g$, are assumed to 
satisfy the following inequality
\be
g^2 \gg \lambda .
\ee
It is a natural assumption since the quartic coupling of the inflaton must be
very small $\lambda \sim 10^{-15}$ while the (gauge) coupling $g$ is not necessarily so.

The equations of motion of the scalars  are given by
\be
&& \square{\phi}   + V'(\phi) + g^2 \chi^2 \phi =0,  \n
&& \square{\chi}  +  g^2 \phi^2  \chi =0.
\ee
The d'Alembertian operator acting on modes with comoving momenta $k$
is given by 
$\square  = \partial_t^2+ 3H  \partial_t + (k^2/a^2),$
where $a$ is the scale factor and the Hubble constant 
is given by $H=\dot{a}/a$,
\be
H^2= \frac{1}{3 M_{\rm pl}^2} 
\left[
\frac{\dot{\phi}^2}{2} + V(\phi, \chi) 
+\frac{\dot{\chi}^2}{2} 
\right]  \   .
\ee
We divide the inflaton field into the coherent motion (zero mode) $\phi_{0}$
and fluctuation (non-zero modes) $\varphi$ as
\be
\phi (t,x) = \phi_{0} (t)+ \varphi (t,x) \    .
\ee
For a sufficiently large initial value $\phi_0 >M_{\rm pl}$, the coherent motion 
of the inflaton field realizes the LFI. 
 The LFI ends  around  $\phi_0 \sim \sqrt{12} M_{\rm pl}$ where $\eta \sim 1$, and starts
falling down towards the minimum. 
The energy density of the universe is dominated by the coherent
oscillation, and the field starts oscillation with 
the effective frequency $\omega_{\rm eff} \sim \sqrt{\lambda} \Phi_0$ 
\footnote{In the quartic potential of $\lambda \phi^4/4$, the solution to the equation of motion $\ddot{\phi_0}+ \lambda \phi_0^3=0$ is given by $\phi_0(t)= \Phi_0  \ {\rm cn} (\sqrt{\lambda} \Phi_0 t ,1/2).$
The Jacobi elliptic function is well approximated by the trigonometric function 
$\phi_0(t) \sim \Phi_0 \cos (\omega_{\rm eff} t)$ where $\omega_{\rm eff}=0.8472 \sqrt{\lambda} \Phi_0$. } .
Here $\Phi_0$ is the amplitude of the oscillation.  
The effective frequency of the inflaton oscillation $\omega_{\rm eff}$ is
 larger than the Hubble constant $H$ after the oscillation starts. Hence,
 the expansion of the universe can be neglected during each
 oscillation of the $\phi$ field. 
It is also worth mentioning here that $\omega_{\rm eff}$ largely changes 
 in finite density state. It is discussed in the next section.

In the coherent motion of the inflaton field, the $\chi$  and the non-zero
modes of $\phi$ acquire time-dependent mass terms:
\be
m_{\rm \chi}(t)^2 = g^2 \phi_0 (t)^2, \ \ \ m_{\rm \varphi}(t)^2 =3 \lambda \phi_0 (t)^2 \  ,
\label{zerodensitymass}
\ee
and, if the adiabaticity  condition  ($|\dot{\omega}|/\omega^2 <1$) is violated,
particle creations by parametric resonance occur. 
It is called the preheating mechanism \cite{preheating}.
The equation of motion for the mode  with momentum $k$ is approximated by
\be
\ddot{\chi}_k + 3 H \dot{\chi}_k + \omega_\chi^2(k)  \chi_k  \sim 0 \ \ 
\mbox{where} \  \ 
\omega_\chi^2(k)= \left( g^2 \Phi_0^2 \sin^2 (\omega_{\rm eff} t) + \delta m^2 +\frac{k^2}{a^2} \right) .
\label{eom-chi1}
\ee
For later convenience we added the term $\delta m^2$ which
represents  slowly changing mass shifts  due to back reactions of the created particles. 
It is a finite density effect and 
 absent until  particles are created by  preheating. This term
plays an important role in section \ref{sec:backreaction}.

As mentioned, the expansion of the universe is much slower than the inflaton oscillation.
Then, by neglecting the Hubble term,
 the equation becomes identical with the Mathieu equation.
The violation of the adiabaticity condition is efficient near $\phi_0 \sim 0$ and for 
smaller $k$. The equation is transformed into the standard form of the Mathieu equation
by defining the new coordinate $z =\omega_{\rm eff} t$,
\be
&& \frac{\partial^2 \chi_k}{\partial z^2} + ( A - 2q \cos 2z  ) \chi_k =0, \n 
&& A=\frac{1}{\omega_{\rm eff}^2} \left( \frac{k^2}{a^2} + \delta m^2 +\frac{(g \Phi_0)^2}{2} \right), \ \ 
q =\frac{|m_\chi|^2}{4 \omega_{\rm eff}^2} = \frac{(g \Phi_0)^2}{4 \omega_{\rm eff}^2}  \   .
\ee
where we defined $|m_\chi|= g \Phi_0$.
In the standard form, $\sqrt{A}$ represents the  ratio of the (averaged) frequency of the 
$\chi_k$ field to the frequency of the external force. On the other hand, $q$ represents the strength 
of the external force. The Mathieu equation describes the phenomena of parametric 
resonances, and the solution is unstable in some regions of $(A,q)$ parameter space.
For $q<1$, the solution is unstable only when $A$ satisfies special conditions for the
narrow resonance \cite{preheating}.
On the contrary, if the external force is sufficiently strong $q>1$, the solution is 
unstable for wide range of parameter space and the ratio of the 
frequencies $A$ is not strongly constrained. This is called the broad parametric resonance. 
In a realistic model, we need to take into account back reactions from the produced particles and the red-shift of momenta by the expansion of the universe. 
They change the narrow resonance condition for $A$,
and hence, in order to realize the rapid increase of 
particles due to the Bose enhancement, the broad resonance condition $q \gg 1$ is 
necessary.  

The particle creation is most efficient around $\phi_0 =0$, and in order to 
estimate the particle creation rate, we  expand
the external force around  $\phi_0 (t)=0.$
The equation of motion (\ref{eom-chi1}) is then approximated by
\be
\ddot{\chi}_k + \left( \frac{k^2}{a^2} + \delta m^2 + |\dot\omega_\chi|^2 t^2 \right) \chi_k =0 \ .
\ee
where $|\dot\omega_\chi| \equiv g \Phi_0 \omega_{\rm eff}$.
It is identified with the Schr\"odinger equation in an inverted harmonic potential, 
and the particle creation due to the Bogoliubov transformation around $t=0$ is
calculated as the tunneling rate in the potential.
The Bogoliubov coefficient $\beta_k$  is given by 
\be
|\beta_k|^2 = e^{-\pi \kappa_k^2}, \ \ \ \kappa_k^2 = \frac{1}{|\dot\omega_\chi|} \left( \frac{k^2}{a^2} + \delta m^2 \right)\ . \label{kappa0}
\ee
A necessary condition for the preheating is 
$\kappa_k \lesssim 1$.  Otherwise, the adiabaticity condition is 
not violated and the particle creations do not occur efficiently. 
This gives an upper bound of momenta of created particles.

To summarize, the particle creation due to the preheating (broad resonance) occurs
only when the two conditions, $q \gg 1$ and $\kappa \lesssim 1$, are satisfied.
For the sufficient particle creation of $\chi$ particles, it is necessary to satisfy
\be
&\cdot& \ \ \ q_\chi = \frac{|m_\chi|^2}{4 \omega_{\rm eff}^2} =
\frac{(g \Phi_0)^2}{4 \omega_{\rm eff}^2} 
\sim \frac{g^2}{ \lambda} \gg 1  
\label{cond-q}
\\
&\cdot& \ \ \ \frac{k^2}{a^2} + \delta m^2  \lesssim  |\dot\omega_\chi|= g\Phi_0 \omega_{\rm eff}  
\label{cond-kappa}
\ee
Here we explicitly write the subscript $\chi$ in $q_\chi$ 
to distinguish it from the same parameter for other particles. 
When these conditions are satisfied together with the Bose enhancement effect, 
the rapid growth of the $\chi$ particles occur.
Then the particle number density increases exponentially as
$ n_{\chi} \propto e^{2 \mu z}.$ 
The coefficient $\mu$ is  ${\cal O}(0.1)$ \cite{preheating}.
It varies model by model, 
but the detailed value of $\mu$ is not essential in the following discussions. 

If the created particles $\chi$ decay or annihilate into other particles, or  
dilute due to the rapid expansion of the universe,
faster than the creation rate $\mu \omega_{\rm eff}$,  
Bose enhancement effect does not work. 
 Hence the following conditions 
\be
&\cdot& \ \ \  \mu  \omega_{\rm eff} > \Gamma_\chi, \ \ H \ 
\label{decayexpansionrate}
\ee 
need to be taken into account for the exponential growth of particle numbers  to occur.
The Hubble parameter $H = \sqrt{\lambda/12} (\Phi_0^2 / M_{\rm pl}) $ is 
smaller than $\mu \omega_{\rm eff} \sim \mu \sqrt{\lambda} \Phi_0$ for $\Phi_0 \ltsim \sqrt{12} \mu M_{\rm pl}$, 
and we can neglect the effect of the expansion of the universe in the period of rapid oscillation.
In the $B-L$ model \cite{IO}, $\chi$ decays into SM particles with the coupling $g$.
Hence the decay width is given by $\Gamma_{\chi} \sim g^2 m_\chi \sim g^3 \Phi_0$
while  the effective frequency is $\omega_{\rm eff}\sim \sqrt{\lambda} \Phi_0.$ 
Hence, if $\mu \sqrt{\lambda}>g^3$, the condition $\mu  \omega_{\rm eff} > \Gamma_\chi$
 is satisfied.
Indeed, as mentioned at the end of section \ref{sec:dynamicaltune}, 
$\lambda \sim g^4 \ll 1$ is required in the model \cite{IO} and thus the decay rate is smaller
than the production rate.

Let's go back to the conditions (\ref{cond-q}) and  (\ref{cond-kappa}) .
The first condition (\ref{cond-q}) is satisfied in our model since $\lambda$ is extremely small. 
When $\delta m^2=0$, the second condition (\ref{cond-kappa}) gives an upper bound of the physical momenta 
$p=k/a$ of the created particles,
\be
p^2 < p_{\star}^2  \ 
\label{maxp}
\ee
where
\be
p_{\star}^2 \equiv |\dot{\omega}_{\chi}| = g \Phi_0 \omega_{\rm eff} 
\sim (g \Phi_{0})^2 /\sqrt{q_{\chi}} \label{p_star}    
\  .
\ee
Since particles with lower momenta are more efficiently produced, 
the  number distribution is far from being in thermal equilibrium. 
For $q_{\chi}\gg1$,  the mass squared $m^2_{\chi}(t)=(g\phi_{0}(t))^2$ is larger than 
$p^2_{\star}$ and the created $\chi$ particles behave non-relativistically, 
except for short intervals of the oscillation satisfying
$\phi_{0}(t) < q_{\chi}^{-1/4}  g \Phi_{0} $. 
In finite density state discussed in the next section,
 the l.h.s. of (\ref{maxp}) is replaced by 
$p^2+\delta m^2$. Therefore, when back reactions of created particles 
generate larger mass corrections $\delta m^2 > p_{\star}^2$,  
the wave equations for $\chi$ with any low momenta behave adiabatically 
 and the particle creations stop.

On the contrary to $\chi$,  $\varphi$ particles (non-zero modes of $\phi$)
are not efficiently produced  by the preheating since 
the  first condition $q \gg 1$  is not satisfied. The parameter $q$ for $\varphi$ is
given by
\be
q_\varphi =  \frac{|m_\varphi|^2}{4 \omega_{\rm eff}^2} 
=\frac{3 \lambda \Phi_0^2}{ 4 \omega_{\rm eff}^2} \sim {\cal O}(1)   \  
\ee
and it does not satisfy the broad resonance condition.
Furthermore,
as seen in the next section, production of $\chi$ particles leads to 
a larger frequency $\omega_{\rm eff}$, and consequently $q_{\varphi}$
becomes much smaller than ${\cal O}(1)$. Hence the preheating never occurs
for $\varphi$.  But instead, 
 $\varphi$ particles can be   rapidly produced   from the $\chi$ particles by 
rescatterings  $\chi + \phi_0 \rightarrow \chi +\varphi$ and annihilations 
$2 \chi \rightarrow \phi_0 + \varphi$, or $2\chi \rightarrow \varphi + \varphi$. 
These particles have momenta $p \lesssim p_\star$.
Since its mass is generated by  finite density effects only, 
 $\varphi$ behaves relativistically \cite{FK}.  It does hold even after
  the finite density effect 
 gives a nonvanishing mass $m_\varphi^2 =\lambda \langle \varphi^2 \rangle$
 because $\lambda \ll g^2$ is assumed in the present analysis.


\section{Back reactions  and the end of preheating}
\label{sec:backreaction}
Once the oscillation of the inflaton starts, the number density of $\chi$ particles 
grow exponentially. As discussed,
 $\mu \omega_{\rm eff}$ is much larger than the Hubble
constant for $\Phi_0 \ll M_{\rm pl}$, and
 the production rate of $n_\chi$ is  always larger
 than the expansion rate of the universe and a precise value of $\mu$ is not important. 
 
Soon after  $n_{\chi}$ increases, the scattering and annihilation process through
the interaction term $g^2 \chi^2 \phi^2$ rapidly create non-zero modes $\varphi$,
and the universe is filled with 
$\chi$ and $\varphi$  particles\footnote{ In the paper \cite{FK}, the evolution of occupation numbers
 were numerically investigated. FIG.13 in \cite{FK}
shows  the rapid growth of $\varphi$  after $\chi$ particles are created by the 
preheating. It also shows that the occupation numbers of $\varphi$ and $\chi$ 
become equal when the exponential growth stops. }. The number density $n_\varphi$
grows until it becomes equal to $n_\chi$ \cite{FK}. 
The $\chi$  field acquires an additional mass correction in addition to (\ref{zerodensitymass}),
\be
&& m_\chi^2 =  g^2(\phi_0^2(t)  + \langle \varphi^2 \rangle) \ .
\label{omega-chi-plasma}
\ee
At the beginning of preheating, the coherent part is dominant, but
soon the second term becomes comparable to the first term.
The  2-point function at the same space-time point
can be evaluated by using the Hartree approximation as
\be
\langle \varphi^2  \rangle = \int \frac{d^3 k}{(2\pi a)^3} \frac{n_{\varphi,k}+1/2}{\omega_k} \sim \frac{n_{\varphi}}{p_{\star}}
\label{varphi-2p-function}
\ee
where  $\omega_{k}$ is replaced by the typical momentum $p_{\star}$
of $\varphi.$ 
In the following we see that the preheating stops when both terms become
comparable and indistinguishable.

The coherent motion of  $\phi$ is also modified by the additional
contribution
$ g^2 \langle \chi^2 \rangle \phi^2 /2$ in finite density state.
The 2-point function for $\chi$ field
is similarly evaluated  as
\be
\langle \chi^2  \rangle = \int \frac{d^3 k}{(2\pi a)^3} \frac{n_{\chi,k}+1/2}{\omega_k}
\sim \frac{n_{\chi}}{g|\phi|}. \ \  \ 
\label{2p-function}
\ee
Here, by using the fact that the created particles are 
nonrelativistic\footnote{At later stages of preheating, 
the energy is transferred from IR to UV momenta due to scatterings
and the $\chi$ particles become relativistic. 
Numerical simulations are necessary for further precise evaluations of $\langle \chi^2 \rangle$.
We leave it for future investigations.},
$\omega_k$ is replaced by  
$m_\chi=g |\phi_0(t)|$.
It is justified as long as $|\phi_0|  > \sqrt{\langle \varphi^2 \rangle}$ 
in (\ref{omega-chi-plasma}).
Then the  interaction between the inflaton and the 
$\chi$ field gives the induced potential which is linear \cite{preheating} in $\phi$,
\be
\frac{g^2}{2}  \langle \chi^2 \rangle \phi^2 \sim
\frac{g^2}{2} \frac{n_\chi}{g |\phi|} \phi^2 = \frac{g}{2} n_\chi |\phi|  \  . 
\label{linearpotential}
\ee
Comparing this with the original potential $\lambda \phi^4/2$,
the above term becomes more dominant when the number of created particles $n_\chi$
is larger than the number density 
determined by the amplitude of the inflaton oscillation; $ n_\chi > (\lambda/g) \Phi_0^3$. 
Hence, the quartic potential of the inflaton is gradually modified by the 
back reaction of created particles, and when the above condition is satisfied,
inflaton  oscillation can be approximated by the linear-type potential $V \propto |\phi|$ for $|\phi| > \sqrt{\langle \varphi^2 \rangle}$.
 The effective frequency of the inflaton oscillation in the linear potential 
  is given by $\omega_{\rm eff}^2 \sim g n_{\chi}/\Phi_0 \sim g^2 \langle \chi^2 \rangle_{\phi_0 = \Phi_0}$.

When the  2-point function $\langle \varphi^2 \rangle$ in (\ref{omega-chi-plasma}) increases up to 
 $\langle \varphi^2 \rangle=\Phi_0^2$, $\langle \chi^2 \rangle$  is 
 replaced by
\be
\langle \chi^2 \rangle = \frac{n_\chi}{g\Phi_0} \ 
\label{2p-function-2}
\ee 
and the inflaton potential changes from the linear to the
 quadratic one for $\phi < \Phi_0$,\footnote{ \label{footnote:1storder}
 In \cite{Felder:2000sf}, from 
 the validity of the quadratic potential for $\phi > \Phi_0$  
and numerical simulations,  the authors concluded  that 
 the effective potential is linear for large $\phi > \Phi_0$ 
 even after the $\varphi$ production up to $\langle \varphi^2 \rangle \sim \Phi_0^2$,
and consequently  the first order phase transition occurs.   } 
 \be
\frac{g^2}{2}  \langle \chi^2 \rangle \phi^2 \sim
\frac{g n_\chi}{ 2 \Phi_0} \phi^2   \  . 
\label{lineartoquadraticpotential}
\ee
Hence the effective frequency of the inflaton oscillation does not changed, and is given by
 $\omega_{\rm eff}^2 = g n_{\chi}/\Phi_0.$

Through the preheating, 
the energy of the coherent motion of inflaton is transferred 
to the energy of the created particles. 
The preheating finally ends either the condition for the broad
parametric resonance $q_\chi>1$ (the value of $q_\chi$ in zero density state
is given in (\ref{cond-q}))  or the condition for the mass correction $\delta m^2<p_\star^2$
is violated. Since $q_\chi, \delta m, p_\star$ are functions of $g, \lambda$,
$\langle \chi^2 \rangle$ and $\langle \varphi^2 \rangle$, we need to numerically solve the 
evolution of 2-point functions $\langle \chi^2 \rangle$, $\langle \varphi^2 \rangle$
and the coherent mode $\Phi_0$.
In the following we evaluate the conditions for broad parametric resonance
based on the above ansatz for the 2-point functions.

The parameter $q_\chi$ gets  modifications 
by the created particles as follows.
The mass of the  $\chi$ field in finite density state
is given by eq.(\ref{omega-chi-plasma}).
On the other hand, the effective frequency $\omega_{\rm eff}$ 
of the inflaton field is given as
\be
\omega_{\rm eff}^2 \sim  \lambda (\Phi_0^2 +  \langle \varphi^2 \rangle) + 
g^2 \langle \chi^2 \rangle  \  ,
\ee
and it receives larger corrections than the bare term $\lambda \Phi_0^2$ 
because $\lambda \ll g^2.$
Accordingly the parameter $q_\chi$ to determine the broad parametric 
resonance for $\chi$ 
particles in the finite density state is replaced by
\be
q_\chi 
=\frac{m_\chi^2}{4\omega_{\rm eff}^2} \sim
\frac{g^2 (\Phi_0^2  + \langle \varphi^2 \rangle)}{\lambda (\Phi_0^2 +  \langle \varphi^2 \rangle) + 
g^2 \langle \chi^2 \rangle } \ .
\ee
As we saw, when created particles are absent, it is $q_\chi=g^2/\lambda \gg 1.$
As the $\chi$ particles are created by preheating (remember $g^2 \gg \lambda$),
the last term becomes  to dominate the denominator and
$\omega_{\rm eff}^2$ is approximated by  $g^2  \langle \chi^2 \rangle$.
The preheating stops when 
$\langle \chi^2 \rangle  \sim \Phi_0^2$ and $q_\chi$ decreases down to ${\cal O}(1)$. 
At the time, the mass  of $\varphi$ becomes 
\be
m_\varphi =\omega_{\rm eff} \sim g \sqrt{\langle \chi^2 \rangle} \sim g \Phi_0 \  .
\ee
Thus,  since $p_{\star} = \sqrt{g\Phi_0 \omega_{\rm eff} }$, 
we have the following relation
\be
m_\chi \sim m_\varphi \sim p_{\star} \sim g \Phi_0  \  
\label{massrelation}
\ee
when the preheating stops.  
If $\varphi$ are rapidly produced  by scatterings, we have
the chemical equilibrium  $n_\varphi \sim n_\chi$. Then
the relation $\langle \varphi^2 \rangle =\langle \chi^2 \rangle=\Phi_0^2$ follows
 (\ref{varphi-2p-function}) and (\ref{2p-function-2}).
This is the case we study in our paper. (See also the footnote \ref{selfinteraction}.)

When the condition $q_\chi>1$ is violated, 
the other condition $\delta m^2 < p_\star^2$ for the preheating
becomes simultaneously violated
in the above situations.
Both sides are given by  $\delta m^2 =g^2 \langle \varphi^2 \rangle $
and $p_\star^2=g\Phi_0 \omega_{\rm eff} \sim g^2 \Phi_0 \sqrt{\langle \chi^2 \rangle}
\sim g^2 \Phi_0^2.$  Hence if $\varphi$ particles are rapidly produced
and  $\langle \varphi^2 \rangle =\langle \chi^2 \rangle=\Phi_0^2$ holds, 
$\delta m^2 = p_\star^2$ is satisfied 
 and even zero-momentum particles cannot be created.

To summarize, all the fluctuations have the same amplitude
$\langle \varphi^2 \rangle =\langle \chi^2 \rangle = \Phi_0^2$ when the preheating stops. 
And the number densities are related to the amplitude of fluctuations as
\be
n_\chi(t) = n_\varphi(t) = g \Phi_0^3 (t) \ .
\ee
Initially, $g \Phi_0^3$ is larger than $n_\chi$ and $n_\varphi$, and the preheating occurs. 
The  amplitude of the coherent oscillation $\Phi$ is reduced and 
the number density $n_\chi$ increases. 
At the same time scatterings  produce $\varphi$ particles 
and increase $n_\varphi$. 
Finally, 
the preheating stops when the number densities become equal to $g \Phi_0^3$.
In addition, the effective masses of the particles become equal $m_{\chi}=m_{\varphi}=g \Phi_0$.
These properties are  consequences of the hierarchy in the coupling constants $g^2 \gg \lambda$.

The initial value  $\Phi_{\rm 0, start}$ of  the coherent motion of the inflaton field 
determines the amplitude  $\Phi_{\rm 0, end}$ at which
the broad parametric resonance stops. 
Since the  particle production occurs faster than the Hubble time scale,
the energy conservation gives a  relation between them as follows. 
The potential energy of the initial inflaton configuration
$\lambda \Phi_{\rm 0, start}^4$ is transfered 
to the energy of the $\chi$ and $\phi$ particles
$\sim g^2 \langle \varphi^2 \rangle \langle \chi^2 \rangle=g^2 \Phi_{\rm 0, end}^4.$
Thus we have
\be
\Phi_{\rm 0, end} = \left(\frac{\lambda}{g^2} \right)^{1/4} \Phi_{\rm 0, start} \  .
\label{phiend}
\ee
Note that the finite density state is not in the thermal equilibrium since the 
preheating is mostly efficient for very low momentum particles.
Further studies of the process towards 
thermal equilibrium  (i.e. turbulence flow from IR to UV momenta)
are left for future investigations \cite{turbulence} \footnote{\label{selfinteraction}
In \cite{turbulence},
the authors studied a case where the created particles $\chi$ interact with themselves
and its number density generates the mass for $\chi$ itself. In such cases, the 
 exponential growth
stops much earlier and the stationary turbulence ( which leads to linear growth 
of occupation numbers ) is shown to occur. In the process, the energy flows from IR
modes to UV modes.}.

After the preheating stops, the energy density is 
dominated by the $\chi$ and $\phi$ particles, and given by
$\rho \sim g^2 \Phi_0^4(t)$. 
In the following, we use $\Phi$ to indicate the typical amplitude of the fluctuations.
As the universe expands, 
the particle densities gradually dilute and the amplitude $\Phi(t)$ decays as well.

\section{Trapping the inflaton near the top of the potential}
\label{sec:trap}
The inflaton potential is modified due to the created particles
and  the potential is raised, which may prevent the inflaton field 
from falling down to the minimum $\phi=M$ of the Coleman-Weinberg 
potential $V(\phi)$.  
Since the magnitude of the  potential raise depends on the amplitude of 
the fluctuations $\langle \chi^2 \rangle=\Phi^2$, 
it is necessary to check whether the field is trapped without falling down to the minimum. 
In this section, we show that it actually happens for wide range of parameters.

\begin{figure}[t]
 \begin{center}
\includegraphics[width=0.8 \linewidth,bb=0 0 720 541]{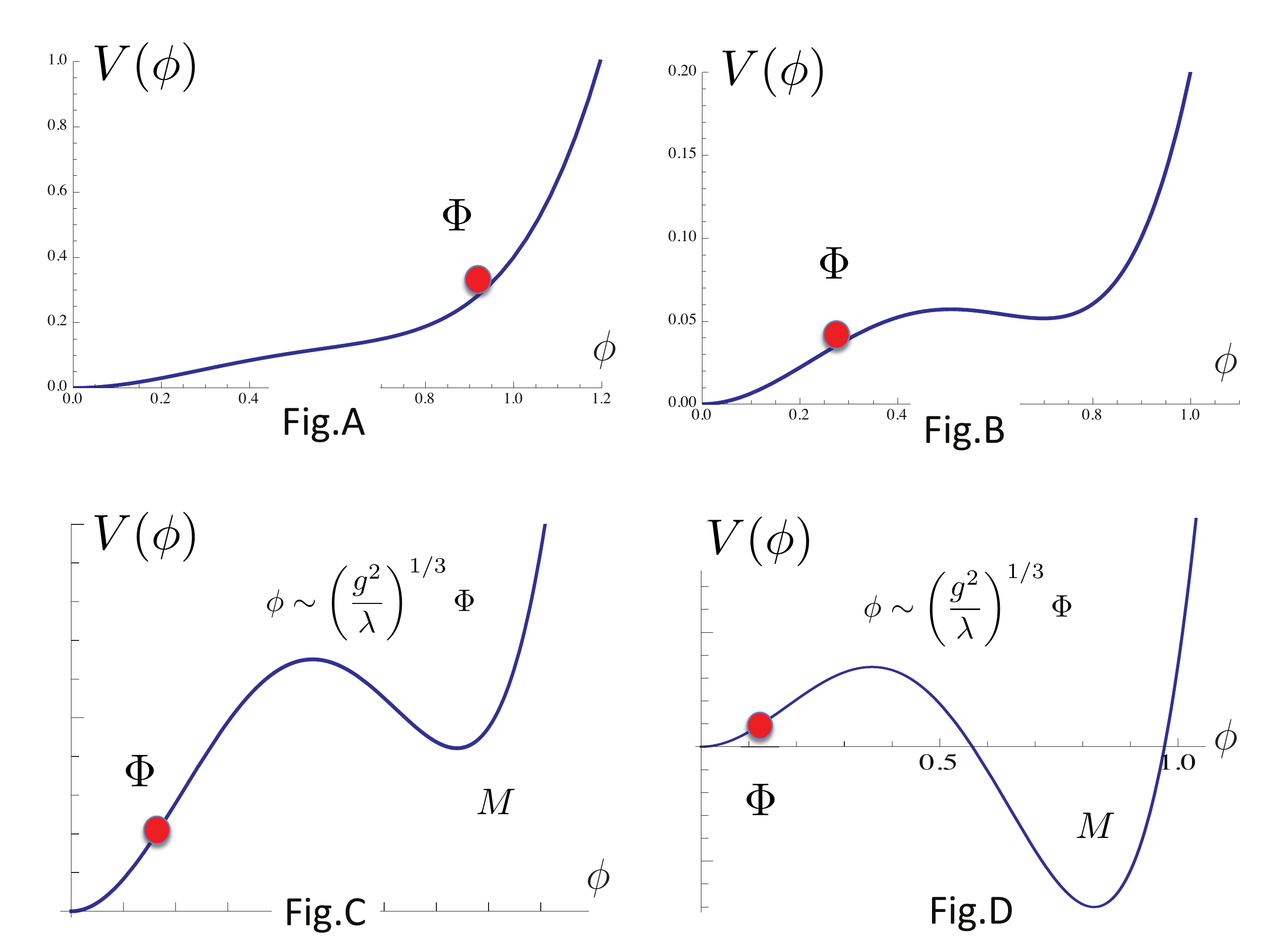}
   \caption{ Schematic figures of the inflaton potential modified by the finite density effect. When the amplitude of the fluctuations $\Phi$ is larger than $\Phi_1 \equiv (4e\lambda/3 g^2)^{1/3} M$, 
   the potential has no nontrivial minimum (Fig.A). As the amplitude becomes smaller,
   the nontrivial minimum appears (Fig.B, C). For smaller amplitude of $\Phi$, 
   the minimum at $\phi=M$   becomes the true minimum as in Fig.D. The position of
   the potential barrier is located around $(g^2/\lambda)^{1/3} \Phi$ and 
   it is always on the right of $\Phi$, if $g^2 >\lambda.$
 }
   \label{fig:Vm}
 \end{center}
\end{figure}

As we saw in the previous section, 
when the number density of created particles becomes comparable to $g \Phi_0^3$,
the preheating stops. 
After the preheating stops, the amplitudes of fluctuations keep the relation
$\langle \chi^2 \rangle=\langle \varphi^2 \rangle=\Phi^2$ since 
$\chi$ and $\phi$ particles
are in chemical equilibrium and the zero mode $\phi_0$
is  indistinguishable from the non-zero modes $\varphi$. 
Keeping this relation, the amplitudes decrease as the universe  expands.

When the amplitude of the fluctuations is $\Phi$,
the inflaton potential is approximated by
\be
V &=& \frac{\lambda }{4}   \phi^4 \left( \ln \frac{\phi^2}{M^2} -\frac{1}{2} \right)
+  \delta V    , \ \ \ 
\delta V = \left\{ \begin{array}{ll}
\frac{g^2 \Phi^2}{2} \phi^2 & |\phi| <  \Phi \\
\frac{g^2 \Phi^3}{2} |\phi| & |\phi| > \Phi
\end{array}\right.
\label{modifiedV}
\ee
The second term $\delta V$ is the finite density contribution of created particles.
The first term also receives an additional  contribution in finite density state, 
but since $g^2 > \lambda$,
it is negligible compared to the last term.
The main role of the first term is to give the minimum at $\phi=M.$
Due to $\delta V$, the minimum at $\phi=M$ is lifted as in Fig.\ref{fig:Vm}.
The question is whether the position of the barrier between two minima
is on the right of the field value $\phi=\Phi.$
For $g^2 > \lambda$, it can be shown that 
the modified potential of (\ref{modifiedV}) has a nontrivial minimum only
when $\Phi \lesssim  \Phi_1 \equiv (4e\lambda/3 g^2)^{1/3} M$. 
Otherwise, the potential has the
unique minimum at $\phi=0$  as in Figure \ref{fig:Vm}A. 
Thus the coherent motion falls towards the origin until the amplitude becomes smaller 
than $\Phi = \Phi_1$. 
For $\Phi < \Phi_1$  in 
 Figure \ref{fig:Vm}B, C, D, the potential has two local minima.
It can be easily shown that  
the position of the barrier between two minima is given by 
$\phi \sim (g^2/\lambda)^{1/3} \Phi.$
Hence, for $g^2>\lambda$, 
the barrier is always on the right of $\Phi$. 
Even after the amplitude decreases further and the finite density effect
is no longer capable to lift the minimum at $\phi=M$ above $V=0$ (see
Fig.\ref{fig:Vm}D ),  
the inflaton field is kept being trapped within the barrier.

The above trapping mechanism occurs only when sufficient particle creation
has finished before the field falls down to the true minimum. 
Namely, in order to trap the field in the potential barrier, 
the quadratic potential must be generated before the field falls down. 
A sufficient condition for this is that, 
 the amplitude of the inflaton  $\Phi_{\rm 0, end}$ at the end of the preheating
 is larger than $M$; $\Phi_{\rm 0,end} > M$.
Therefore, using (\ref{phiend}), if $\Phi_{\rm 0, start}$ satisfies the condition
\be
\Phi_{\rm 0,start} >(g^2/\lambda)^{1/4} M    \  ,
\ee
$\chi$ particles are sufficiently produced so that it can
 trap the inflaton field  within the potential barrier.\footnote{To get the actual lower bound on $\Phi_{\rm 0, start}$, accordingly the field value where the zero mode $\phi_{0}$ starts rolling down, we need a detailed numerical calculation with being careful of the shape of the Colman-Weinberg potential (\ref{CWpotential}).
It's beyond the scope of this paper and left as a future work.
Depending on the initial value, new inflation with a sufficient e-folding could occur without the inflaton oscillation \cite{Yokoyama}.}
The field value $\Phi_{\rm 0, start}$ can be much 
smaller than the Planck scale, and the  beyond-Planck-scale problem is absent.

Finally  we comment on the effects of  thermalization of $\chi$ particles
on the trapping mechanism. In particular, we take the $B-L$ model as an example.
 When the preheating stops, the relation $n_\chi =g \Phi^3$ holds. 
In the $B-L$ model, as discussed after eq.(\ref{decayexpansionrate}),
the decay  rate of $\chi$ is given by $\Gamma_\chi=g^3 \Phi.$
$\chi$ can also annihilate into the SM particles, whose  rate is estimated as
$\Gamma_{\chi, annih} \sim g^4 n_\chi /m_\chi^2 \sim g^3 \Phi.$ 
Hence, if $\Phi <g^2 M_{\rm pl}$, 
 these rates are larger than the Hubble parameter $H \sim g \Phi^2/M_{\rm pl}$
and the thermal bath with the temperature $T \sim \sqrt{g} \Phi$ is produced.
Furthermore, when $\Phi <g^{7/2} M_{\rm pl}$,
 the $B-L$ scattering rate satisfies $\Gamma_{scatt} \sim g^4 T>H$
and all the system including the $B-L$  and the inflaton fields is thermalized.
Then the fluctuation of fields are determined by the temperature 
$\Phi_T^2 \equiv \langle \varphi^2 \rangle \sim \langle \chi^2 \rangle \sim T^2 \sim g \Phi^2$. 
They are smaller than the original value of the fluctuations
 $\langle \varphi^2 \rangle \sim \langle \chi^2 \rangle = \Phi^2$.
This is because thermalization  transfers  energy   from IR to UV regions with typical momentum $p\sim T$
and reduces the large fluctuation produced by the preheating. 
Consequently the coefficient of the quadratic term in (\ref{modifiedV})
is replaced by $g^2 \langle \chi^2 \rangle = g^2 \Phi_T^2$ for $|\phi|<\Phi_T / g$.
In this case, the position of the barrier is given by $(g^2/\lambda)^{1/2}\Phi_T$.  
For the amplitude of the fluctuation is  $\Phi_T$, the barrier is 
always on the right of the fluctuation and the trapping mechanism similarly holds.

\section{Dynamical fine-tuning of the initial condition}
\label{sec:dynamicaltune}
We now determine the initial value of the SFI.
The amplitudes of the oscillation and fluctuations
decrease in the expanding universe where
the energy density  is dominated by the energy of the created particles,
$\rho \sim g^2 \Phi^4.$ 
But as the amplitude becomes  smaller, the vacuum energy $V_0=\lambda M^4/8$ will
dominate the energy of created particles. 
By comparing these energies, we see that the
de Sitter expansion starts when the  amplitude becomes smaller than 
the following value,
\be
\Phi \sim \left( \frac{\lambda}{8 g^2} \right)^{1/4} M  \  .
\label{SFIdeSitter}
\ee
It is already small but still much larger than the necessary 
initial condition in eq.(\ref{initialcondition}). 

During the de Sitter expansion, 
the inflaton  continues to oscillate until the effective frequency of the inflaton 
$\omega_{\rm eff}=g \Phi$ becomes smaller than the Hubble constant 
$H=\sqrt{\lambda M^4/24 M_{\rm pl}^2}$ of the de Sitter universe. 
Hence  the amplitude of fluctuations
continues to decay as far as the condition $\omega_{\rm eff}>H$
is satisfied. The oscillation of the inflaton finally stops when the condition
\be
\Phi \lesssim \frac{1}{g}\sqrt{ \frac{\lambda}{24}} \frac{M^2}{M_{\rm pl}},
\label{phi-dynamicallyset}
\ee
is satisfied. After this condition is satisfied, 
the inequality $\omega_{\rm eff} <H$ holds and 
 the fluctuations of the inflaton field with lower momenta than the Hubble constant
 are {\it frozen}.
Therefore, in the new inflation model with the Coleman-Weinberg potential, 
the amplitudes of the coherent motion and also the fluctuations are reduced
to the very small value\footnote{
If the system is thermalized as discussed at the end of 
section \ref{sec:trap}, the energy density of the universe
$\rho$ is given by $\sim \Phi_T^4$ and the r.h.s.
of (\ref{SFIdeSitter}) is
replaced by $(\lambda /8)^{1/4} M$.
During the de Sitter expansion,
the scattering rate $\Gamma_{scatt}$ become smaller than the Hubble constant $H$.
Then the system is no longer in thermal equilibrium. 
But the amplitude of fluctuation $\Phi_T$ is red-shifted and
continues to reduce as $a^{-1}$. Since the effective frequency 
of inflaton is given by $\omega_{\rm eff}=g \Phi_T$, the fluctuation 
is frozen at the same value of (\ref{phi-dynamicallyset}).
}.

Eq.(\ref{phi-dynamicallyset}) solves the fine-tuning problem of the small field inflation
 (\ref{initialcondition}). 
Let us estimate  the coefficient in (\ref{phi-dynamicallyset}) numerically. 
The quartic coupling is determined by the amplitude of  the curvature fluctuations as
$\lambda \sim 10^{-15}$. Inserting the value in  (\ref{phi-dynamicallyset}), it becomes
\be
\Phi \sim 10^{-3} \left( \frac{10^{-5}}{g} \right) \frac{M^2}{M_{\rm pl}}.
\label{finalanswer}
\ee
It is smaller than the upper bound of (\ref{initialcondition})
if the coupling $g$ satisfies $g \gtrsim10^{-5}$. 
For general models $g$ is a free parameter and we can take
any value, but in the $B-L$ extension of the SM \cite{IO}, the 
$\beta$-function of  the quartic coupling 
$\lambda$ has a contribution from the gauge coupling;
$\beta_\lambda = 96 g^4/16\pi^2$. Hence unless $g^4 \sim \lambda$, 
we need a fine-tuning to keep the smallness of  $\lambda$. 
The most natural assumption is  $g \sim \lambda^{1/4} \sim 10^{-4}.$
Thus, within the model \cite{IO}, the initial condition problem of the SFI is 
naturally solved.


\section{ Summary}
\label{sec:summary} 
In this paper, we proposed a mechanism
to solve the fine-tuning problem of the new inflation, the small field 
inflation with the Coleman-Weinberg type potential. 
The key relation (\ref{phi-dynamicallyset}) to determine the initial value is 
obtained by comparing the effective frequency of the oscillation and the Hubble
constant ($H^2=V_0/3M_{\rm pl}^2$).  
The flatness at the top of the potential is responsible for the fine-tuning problem 
of the SFI. Corresponding to this fact, the effective frequency should be
 dynamically generated
by the fluctuations of created particles.
Then,  from the dimensional analysis,  we can expect that 
it is given by $\omega_{\rm eff}=g \Phi$, where $g$ is the coupling to the field
which gives effective potential of the inflaton.
Then, if  the vacuum energy is given by $V_0=\lambda M^4$, the initial value of the inflaton is given by $\Phi \sim \sqrt{\lambda}/g (M^2/M_{\rm pl})$.  
Therefore, once the inflaton field is trapped
by the quadratic potential generated in the preheating, the fine-tuning
problem in similar models can be solved. 

For hilltop inflation models with a
negative curvature potential ($-\mu^2 \phi^2/2$) at the origin, 
the dynamical fine-tuning mechanism for the small field inflation does not work since the system undergoes the first order phase transition before a sufficiently large e-folding number is gained \cite{Felder:2000sf}.
Hence the flatness ($\mu=0$) at the top of the hill is essential to 
solve the fine-tuning problem of the initial condition in SFI. 
It is interesting that two completely different  fine-tuning problems,
the Higgs mass \cite{naturalness} and the initial condition in SFI, can be solved dynamically
by simply assuming the absence of the dimensionful parameter $\mu$
in the bare Lagrangian.

\section*{Acknowledgments}
We would like to thank Hideo Kodama  for 
useful discussions and comments.
This work is supported by the Grant-in-Aid for Scientific research from the
Ministry of Education, Science, Sports, and Culture, Japan,
Nos. 23540329, 23244057 (S.I.), 26105520, 26247042, 15H05889 (K.K.).  

\end{document}